\documentclass[12pt]{article}
\usepackage{latexsym}
\usepackage{epsfig}
\usepackage{amsmath}

\begin{document}
\vspace*{-.6in}
\thispagestyle{empty}
\begin{flushright}
CALT-68-2502
\end{flushright}
\baselineskip = 18pt

\vspace{1.5in} {\Large
\begin{center}
THE GENERALIZED STIELTJES TRANSFORM \\
AND ITS INVERSE\end{center}} \vspace{.5in}

\begin{center}
John H. Schwarz
\\
\emph{California Institute of Technology\\ Pasadena, CA  91125, USA}
\end{center}
\vspace{1in}

\begin{center}
\textbf{Abstract}
\end{center}
\begin{quotation}
\noindent The generalized Stieltjes transform (GST) is an integral
transform that depends on a parameter $\rho > 0$. In previous work
a convenient form of the inverse transformation was derived for
the case $\rho = 3/2$. This paper generalizes that result to all
$\rho > 0$. It is a well-known fact that the GST can be formulated
as an iterated Laplace transform, and that therefore its inverse
can be expressed as an iterated inverse Laplace transform. The
form of the inverse transform derived here is a one-dimensional
integral that is considerably simpler.
\end{quotation}

\newpage

\pagenumbering{arabic}

\section{Introduction}

In \cite{He:2002zu} we encountered the integral equation
\begin{equation} \label{inteqn}
 g(\zeta) = \int_0^\infty (\zeta^2 + \mu^2)^{-3/2} f(\mu) \, d\mu
 \quad |\arg\zeta| < \pi/2,
\end{equation}
where $ g(\zeta)$ was a known function, and we needed to solve for
$f(\mu)$. That paper proved that if $g$ has suitable analyticity
and asymptotic properties, which were satisfied for the specific
function of interest, then
\begin{equation}\label{fsoln}
f(\mu) = - \frac{i\mu^2}{\pi} \int_0^\pi {\rm cos}\, \theta ~
g(-i\mu {\rm cos}\, \theta)\, d\theta .
\end{equation}
This result is reviewed in \cite{Schwarz:2003zf}.

By a change of variables, eq.~(\ref{fsoln}) can be rewritten in
the form
\begin{equation}\label{fsoln2}
f(\mu) = \frac{\mu}{\pi i} \int_{-i\mu}^{i\mu}
\frac{\zeta}{\sqrt{\mu^2 +\zeta^2}} ~ g(\zeta)\, d\zeta .
\end{equation}
Then letting $y=\mu^2$ and $z = \zeta^2$, as well as $F(y) =
\frac{1}{2\mu} f(\mu)$ and $G(z) = g(\zeta)$, we obtain
\begin{equation} \label{Geqn0}
 G(z) = \int_0^\infty (y+z)^{-3/2} F(y) \, dy, \quad | \arg z| < \pi
\end{equation}
and
\begin{equation} \label{Feqn0}
F(y) = \frac{1}{4\pi i}  \,  \int_{{\cal C}_y} \frac{G(z)}{\sqrt{
y+z}} dz.
\end{equation}
The contour ${\cal C}_y$ starts and ends at $-y$ and encloses the
origin in the counterclockwise sense.

It is convenient to have a $y$-independent contour, so letting
$z=wy$ we obtain
\begin{equation} \label{Feqn02}
F(y) =  \frac{1}{4\pi i}  \, \sqrt{y} \int_{\cal C}  \,
\frac{G(yw)}{\sqrt{1+w}} \, dw,
\end{equation}
where ${\cal C}$ is a contour that starts and ends at the point $w
= -1$ enclosing the origin in the counterclockwise sense. It could
be chosen to be the unit circle, for instance. Note that it is not
really a closed contour, since $G(z)$ has a branch cut running
along the negative real axis. An integration by parts allows this
to be rewritten in the form
\begin{equation} \label{Feqn03}
F(y) = - \frac{1}{2\pi i}  \, y^{3/2} \int_{\cal C} \sqrt{1+w}\,
G'(yw) \, dw.
\end{equation}

In this paper, we will formulate and prove a one-parameter
generalization of the preceding result. Specifically, we claim
that the integral transform
\begin{equation} \label{Geqn}
 G(z) = \int_0^\infty (y+z)^{-\rho} F(y) \, dy, \quad | \arg z| < \pi
\end{equation}
has as its inverse transform
\begin{equation} \label{Feqn2}
F(y) = - \frac{1}{2\pi i}  \, y^{\rho} \int_{\cal C} (1+w)^{\rho -
1} G'(yw) \, dw.
\end{equation}
This integral converges for $\rho >0$. The special case discussed
above corresponds to $\rho = 3/2$. The counterpart of
eq.~(\ref{Feqn02}), obtained by integration by parts, namely
\begin{equation}
F(y) = \frac{1}{2\pi i} (\rho - 1) \, y^{\rho - 1} \int_{\cal C}
\,(1+w)^{\rho - 2} {G(yw)} \, dw,
\end{equation}
is less general, since it is only valid for $\rho >1$. This is
allowed in the $\rho = 3/2$ case, of course.

We should note for the record what is being assumed about $F$ and
$G$. Specifically, $F(y)$, which is only defined on the positive
real axis, is allowed to be an arbitrary distribution (or
``generalized function"). We require that there exists a number
$\alpha$ with $0< \alpha < \rho$ such that $| \int_{y_1}^{y_2}
y^{\alpha - \rho} F(y)\, dy |$ is bounded by a number independent
of $y_1$ and $y_2$ for all $0< y_1 < y_2$. The function $G(z)$ is
then holomorphic throughout the cut plane $|\arg z | < \pi $, and
there exists a positive real number $\beta$ such that
$|z^{\beta}\, G(z)|$ is bounded at infinity.

The $\rho = 1$ case of eq.~(\ref{Geqn}), known as the Stieltjes
transform, is discussed in Widder's classic treatise on the
Laplace transform \cite{Widder}. The only other cases considered
by Widder are positive integer values of $\rho$, which are related
to the $\rho = 1$ case by differentiation. Following
\cite{Erdelyi}, we refer to the case of arbitrary $\rho$ as the
generalized Stieltjes transform (GST). (In ref.~\cite{Zayed} it is
called a Stieltjes transform of index $\rho$.) The formula for the
inverse GST in eq.~(\ref{Feqn2}) does not seem to have been found
previously. Certainly, it does not appear in \cite{Widder},
\cite{Erdelyi}, \cite{Zayed}, or \cite{Misra}.

Much of the literature on the GST is concerned with the asymptotic
behavior of $G(z)$ for large $|z|$ \cite{Tuan}. We will not
address that topic here. As it happens, in \cite{He:2002zu} we
were interested in deducing the asymptotic behavior of $F$
associated with a given $G$.

In the special case $\rho = 1$ the transform in eq.~(\ref{Geqn})
reduces to the Stieltjes transform
\begin{equation} \label{Hilbert}
G(z) = {\cal S}_z[F] = \int_0^{\infty} \frac{F(y)}{y+z} dy.
\end{equation}
Since the Stieltjes transform is well understood, this case
provides an instructive test of the proposed inverse transform.
Setting $\rho = 1$ in eq.~(\ref{Feqn2}) gives an expression that
can be integrated explicitly to give
\begin{equation}\label{disc}
F(y) = \lim_{\epsilon \to \, 0^+} \frac{1}{2\pi i} \Big(
G(-y-i\epsilon) - G( -y + i \epsilon) \Big).
\end{equation}
Given the stated analytic and asymptotic properties of $G(z)$,
it is a simple consequence of Cauchy's
theorem that this is the correct solution of eq.~(\ref{Hilbert}) for $y>0$.

In order to convince oneself that eqs.~(\ref{Geqn}) and
(\ref{Feqn2}) are plausible for all $\rho >0$,
it is instructive to consider a simple
example. Specifically, if one chooses
\begin{equation} F(y) = y^{\nu -1},
\end{equation}
then the integral in eq.~(\ref{Geqn}) converges for $0 < \nu <
\rho $ and gives
\begin{equation}
G(z) = z^{\nu-\rho} B(\nu, \rho -\nu),
\end{equation}
where $B(u,v)$ is the Euler beta function. It is straightforward
to verify that this pair of functions also satisfies
eq.~(\ref{Feqn2}).

\section{Other versions of the inverse transform}

The change of variables $z=wy$ allows us to rewrite
eq.~(\ref{Feqn2}) in the alternative form
\begin{equation} \label{Feqn3}
F(y) = - \frac{1}{2\pi i}  \,  \int_{{\cal C}_y}
(y+z)^{\rho - 1} G'(z) \, dz,.
\end{equation}
As before, the contour ${\cal C}_y$ starts and ends at the point $z = -y$,
encircling the origin in the counterclockwise sense.
If $\rho > 1$, an integration by parts
brings eq.~(\ref{Feqn3}) to the form
\begin{equation} \label{Feqn}
F(y) = \frac{1}{2\pi i} (\rho - 1) \,  \int_{{\cal C}_y}
(y+z)^{\rho - 2}\, G(z) \, dz .
\end{equation}
Since this is only well-defined for $\rho > 1$,
eqs.~(\ref{Feqn2}) and~(\ref{Feqn3}) are more general than
eq.~(\ref{Feqn}).

Let us define the quantity that appears
on the right-hand side of eq.~(\ref{disc}) to be
\begin{equation}\label{Deltaform}
\Delta(t) = \frac{1}{2\pi i} \lim_{\epsilon \to \, 0^+} \Big(
G(-t-i\epsilon) - G(-t + i \epsilon) \Big), \quad t>0.
\end{equation}
We showed that $F(y) = \Delta(y)$ when $\rho = 1$, but this is not
the case for other values of $\rho$. By shrinking the contour
${\cal C}_y$ down to the cut, eq.~(\ref{Feqn}) takes the form
\begin{equation}\label{special}
F(y) = (\rho - 1) \int_0^y (y-t)^{\rho -2} \Delta(t) \, dt.
\end{equation}
In similar fashion, eq.~(\ref{Feqn3}) gives rise to
\begin{equation} \label{desired}
F(y) =   \int_0^y  (y-t)^{\rho - 1} \Delta'(t) \, dt.
\end{equation}
However, these formulas are only correct if the behavior of $G(t)$
near the origin is such that these integrals exist. The contour
integral versions of these formulas are more general, since they
do not have this restriction.

Equations~(\ref{special}) and (\ref{desired}) have the structure of
Abel transforms. The inverse Abel transform is well-known and can be
used to give a formula for the discontinuity across the cut, $\Delta(t)$,
in terms of the original (generalized) function $F(y)$. A version
that is suitable if $\rho < 2$ and $F(0) = 0$ is
\begin{equation}
\Delta(t) = \frac{\sin \pi \rho}{\pi\, (1-\rho)}
\int_0^t (t-y)^{1-\rho} F'(y) \, dy.
\end{equation}
This gives the known result for $\rho =1$, namely $\Delta(t) = F(t)$.
It can be checked for the simple example $F(y) = y^{\nu -1}$ discussed
earlier.

\section{Proof of the main result}

Let us review how the GST is related to the Laplace transform, for
which we use the following notation
\begin{equation}
\label{Laplace} {\cal L}_x [F] = \int_0^{\infty} e^{-xy} F(y)\,
dy.
\end{equation}
Inserting the identity
\begin{equation}
(y+z)^{-\rho} = \frac {1}{\Gamma (\rho)} \int_0^{\infty} x^{\rho -
1} e^{-x(y+z)} dx
\end{equation}
into eq.~(\ref{Geqn}) gives
\begin{equation} \label{dblLaplace}
G(z)  = \frac {1}{\Gamma (\rho)} \int_0^{\infty} x^{\rho - 1}
e^{-xz} {\cal L}_x [F] \, dx = \frac {1}{\Gamma (\rho)} {\cal L}_z
\big[ x^{\rho - 1} {\cal L}_x [F] \big].
\end{equation}
This recasts eq.~(\ref{Geqn}) as a Laplace transform of a Laplace
transform. In particular, setting $\rho = 1$, this gives the
well-known result that the Stieltjes transform is the square (in
the operator sense) of the Laplace transform, i.e. ${\cal S} =
{\cal L}^2$.

Equation~(\ref{dblLaplace}) implies that
\begin{equation} \label{altG}
{\cal L}_x [F] = \Gamma (\rho) x^{1 -\rho} {\cal L}^{-1}_x [G],
\end{equation}
where ${\cal L}^{-1}$ denotes an inverse Laplace transform. A second
inverse Laplace transform gives
the formal inversion of eq.~(\ref{dblLaplace})
\begin{equation} \label{dblinverse}
F(y) = \Gamma (\rho) {\cal L}^{-1}_y \Big[x^{1 -\rho} {\cal
L}^{-1}_x [G] \Big].
\end{equation}
This can be made very explicit by using the standard contour
integral realization of the inverse Laplace transformation
\cite{Zayed}. The new claim is that eq.~(\ref{dblinverse}) can be
simplified to take the form of eq.~(\ref{Feqn2}).

Let us now carry out some similar manipulations of
eq.~(\ref{Feqn2}). Substituting
\begin{equation}
G'(yw) = -\int_0^{\infty} dt\, e^{-tyw} t {\cal L}^{-1}_t [G]
\end{equation}
into eq.~(\ref{Feqn2}) and taking the Laplace transform of both
sides recasts eq.~(\ref{Feqn2}) in the form
\begin{equation} \label{altF}
 {\cal L}_x [F] = \Gamma (\rho +1) \frac{1}{2 \pi i}
 \int_{\cal C} dw (1+w)^{\rho - 1} \int_0^{\infty} dt
 (x+tw)^{-\rho -1} t {\cal L}^{-1}_t [G].
\end{equation}
Comparing eqs.~(\ref{altG}) and (\ref{altF}), we see that their
equivalence requires the identity
\begin{equation} \label{deltarule2}
 \frac{t \rho}{2 \pi i}
 \int_{\cal C} dw (1+w)^{\rho - 1}
 (x+tw)^{-\rho - 1} = x^{1 -\rho} \delta(x-t),
\end{equation}
or equivalently
\begin{equation} \label{deltarule}
 \frac{\rho}{2 \pi i}
 \int_{\cal C} dw (1+w)^{\rho - 1}
 (x+tw)^{-\rho - 1} = x^{-\rho} \delta(x-t).
\end{equation}

Another way of understanding the necessity of
eq.~(\ref{deltarule}) is to consider the special case $F(y) =
\delta(y-t)$. In this case the GST is $G(z) = (t+z)^{-\rho}$. The
inverse transform eq.~(\ref{Feqn2}) for this choice of $G(z)$
corresponds precisely to eq.~(\ref{deltarule}). This remarkable
equation, which is required to hold for $x,t >0$, is the heart of
the matter. Its proof, which is rather nontrivial, is presented in
the next section.

\section{Representation of a delta function}

This section proves the key formula, namely eq.~(\ref{deltarule}).
Consider the left side of eq.~(\ref{deltarule})
\begin{equation}
\label{deltaone} \chi_{\rho}(x,t) = \frac{\rho}{2 \pi i}
\int_{\cal C} dw (1 + w)^{\rho - 1} (x+tw)^{-\rho-1},
\end{equation}
where $x,t>0$ and ${\cal{C}}$ is the contour shown in fig.~1. By
making the change of variables $w\to 1/w$, it is easy to prove
that $\chi_{\rho}(x,t) = \chi_{\rho}(t,x)$. For $t>x$ the
singularity structure is shown in fig.~1(a), and the contour can
be pushed off to infinity, giving zero for the integral. For
$t<x$, the contour encloses no singularity (see fig.~1(b)), so the
result is again zero, as required by $x\leftrightarrow t$
symmetry.

\begin{figure}[h]
\centering{\psfig{figure=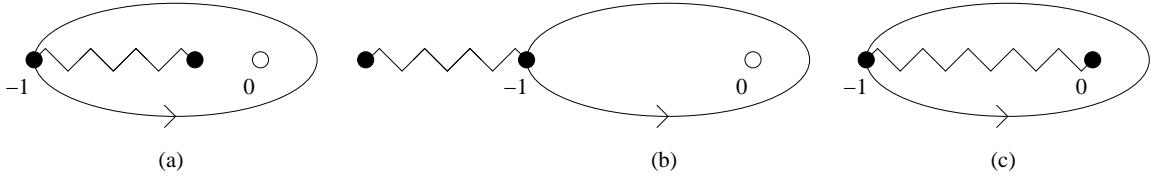,width=6.0in}} \caption{The
contour ${\cal{C}}$ starts and ends at the point $w=-1$ enclosing
the origin in the counterclockwise sense. Three possible branch
cut configurations are depicted.}
\end{figure}

Since $\chi_{\rho}(x,t)$ vanishes for $t\ne x$, it must be some
sort of distribution concentrated at $t=x$. If we assume that it
is proportional to a delta function, i.e. $f(x) \delta(x-t)$, it
is easy to derive $f(x)$. Integrating over $t$ from $0$ to $y$,
gives
\begin{equation}
\label{fderive} f(x) \theta(y-x) = \frac{1}{2 \pi i} \int_{\cal C}
\frac{dw}{w} (1 + w)^{\rho - 1} \Big[x^{ -\rho} - (x+yw)^{
-\rho}\Big].
\end{equation}
The first term on the right is an elementary contour integral and
gives $x^{-\rho}$. The second term is also easily evaluated by
Cauchy's theorem. It gives $-x^{-\rho}$ if $x > y$ and zero if
$y>x$. Thus we deduce that $f(x) = x^{-\rho}$, and hence
\begin{equation}
\label{chiresult} \chi_{\rho}(x,t) = x^{-\rho} \delta(x-t) =
t^{-\rho} \delta(x-t).
\end{equation}

Eq.~(\ref{chiresult}) is the correct result, but the derivation
given above is not rigorous. It assumes that the distribution is
proportional to a delta function, and that it does not involve any
derivatives of delta functions. A more careful analysis that is
sensitive to such terms if they are present involves checking the
proposed answer by integrating both sides against the test
function $e^{-t y}$, i.e., comparing the Laplace transform of both
sides of the equation. Thus we need to show that
\begin{equation}
\label{needtoshow} \int_0^\infty dt \ \chi_{\rho}(x,t) e^{-t y} =
x^{-\rho} e^{-xy}.
\end{equation}

Evaluating the Laplace transform of eq.~(\ref{deltaone}) gives
(after some simple manipulations)
\begin{equation}
\label{remain} \int_0^\infty dt \ \chi_{\rho}(x,t) e^{-t y} =
\frac{\rho}{2 \pi i}\, y^{\rho } \int_{\cal{C}} dw (1 + w)^{\rho -
1} I(xyw),
\end{equation}
where
\begin{equation}
I(u) = e^u \int_u^{\infty} e^{-v} v^{-\rho -1} dv.
\end{equation}
In order to reexpress this function in a more convenient form, we
first note that
\begin{equation}
I'(u) = I(u) - u^{-\rho-1}.
\end{equation}

Let us now assume that $\rho$ is not an integer, which is the case
of most interest. This equation is then solved by an expression of
the form
\begin{equation}
I(u) = I_1 (u) + u^{-\rho} I_2 (u),
\end{equation}
where $I_1 (u)$ and $I_2 (u)$ are regular at $u=0$. The first term
is $I_1(u) = \Gamma (-\rho)\, e^u$. Its contribution to
eq.~(\ref{remain}) is zero, since there is no singularity inside
the contour. The function that matters is $I_2$, which satisfies
the differential equation
\begin{equation}
\label{results} u I_2'(u) - (u+\rho ) I_2 (u) +1 = 0.
\end{equation}
Substituting a power series expansion,
\begin{equation}
\label{I2series} I_2(u) = \sum_{n=0}^{\infty} c_n u^n,
\end{equation}
one obtains the recursion relation $(n+1 - \rho) c_{n+1} = c_n$.
Thus, since $c_0 = I_2(0) = 1/\rho $, we conclude that
\begin{equation}
\label{cnresult} c_n = \frac{\Gamma(1-\rho)}{\rho \, \Gamma(n+1 -
\rho)}.
\end{equation}

To determine the contribution of each term in eq.~(\ref{I2series})
to eq.~(\ref{remain}), we need to evaluate
\begin{equation}
 \frac{1}{2 \pi i}\int_{\cal{C}} dw (1 +
w)^{\rho - 1} w^{n -\rho}  = -\frac{\sin \pi (n -\rho)} {\pi}
B(n-\rho+1, \rho ) = \frac{(-1)^n}{\rho \, c_n\, n!}.
\end{equation}
Combining these results, the $c_n$ factors cancel, and we learn
that
\begin{equation}
\label{proved} \int_0^\infty dt \, \chi_{\rho}(x,t) \, e^{-t y} =
x^{-\rho}\sum_{n=0}^{\infty} \frac{(-xy)^n}{n!} = x^{-\rho}
e^{-xy},
\end{equation}
which is the result we set out to prove. Even though this
derivation needs to be modified when $\rho$ is a positive integer,
the result clearly is valid in that case as well. This completes
the proof of eq.~(\ref{needtoshow}) and hence of the inverse
transform eq.~(\ref{Feqn2}).

\section{An alternative derivation}

In this section we present a simpler, though less general,
derivation of the inverse transform. It follows from Cauchy's
theorem and the required analytic and asymptotic properties of
$G(z)$ that
\begin{equation} \label{Stieltjes}
G(z) =  \int_0^{\infty} \frac{\Delta(t)}{z+t} \, dt,
\end{equation}
where $\Delta(t)$  is defined in eq.~(\ref{Deltaform}). The
validity of this formula requires that $\Delta(t)$ is not too
singular as $t\to 0$, so that the integral exists. This is a
significant restriction, since if $F \sim y^{\nu -1}$ with $\nu
>0$ for small $y$, then $G(z) \sim z^{\nu - \rho}$ for small $z$.
Thus one would need that $\rho < \nu + 1$. No such assumption has
been made previously, which is why the derivation in this section
is less general.

Equation~(\ref{Stieltjes}) is precisely a Stieltjes transform, $G
= {\cal S} [ \Delta]$. We noted earlier that this is an iterated
Laplace transform, ${\cal S} = {\cal L}^2$. Therefore,
\begin{equation}
{\cal L}_x^{-1} [G] = {\cal L}_x[\Delta].
\end{equation}
Comparing this with eq.~(\ref{altG}), which we obtained from
eq.~(\ref{Geqn}), we learn that
\begin{equation}
{\cal L}_x[F] = \Gamma(\rho)\, x^{1-\rho} {\cal L}_x [\Delta].
\end{equation}
Now using ${\cal L}_x[t^{\nu -1}] = \Gamma(\nu) \, x^{-\nu}$, this becomes
\begin{equation}
{\cal L}_x[F] = (\rho -1)\, {\cal L}_x[t^{\rho - 2}]\, {\cal L}_x [\Delta].
\end{equation}
By the convolution theorem, this implies that
\begin{equation}
F(y) = (\rho - 1) \int_0^y (y-t)^{\rho -2} \Delta(t) \, dt,
\end{equation}
which is eq.~(\ref{special}).  The alternative form in
eq.~(\ref{desired}) extends the range of validity for $\rho$, but
is even more restricted in its requirements for the behavior of
$\Delta(t)$ at the origin. As we pointed out in sect.~2, there is
no such issue for the corresponding contour integrals.

\section*{Acknowledgments}
I am grateful to Y. H. He, M. Spradlin, and A. Volovich for their
collaboration in \cite{He:2002zu}, which contains the $\rho=3/2$
case of the result described here. This work was supported in part
by the U.S. Dept. of Energy under Grant No. DE-FG03-92-ER40701.

\newpage

\section*{Appendix: Examples}

In this appendix we list pairs of functions that are related by
the generalized Stieltjes transform and its inverse given in
eqs.~(\ref{Geqn}) and (\ref{Feqn2}). The two simple examples in
the table were discussed in the introduction and in sect.~2. Four
pages of additional examples can be found in Sect. 14.4 of
\cite{Erdelyi}. One of them is given on the last line of the
table. The example in \cite{He:2002zu}, which is specific to $\rho
= 3/2$ (though it can probably be generalized), is not contained
in \cite{Erdelyi}.

\begin{table}[h]
\center{ \begin{tabular}{|c|c|} \hline & \\
$F(y) = - \frac{1}{2\pi i}  \, y^{\rho}
\int_{\cal C} (1+w)^{\rho - 1}  G'(yw) \, dw,
\quad \rho >0$  & $G(z) = \int_0^\infty (y+z)^{-\rho}
F(y) \, dy$  \\ &\\
\hline \hline & \\
$y^{\nu -1}, \quad 0 < \nu < \rho $ &
$B(\nu, \rho - \nu)\, z^{\nu -\rho} $ \\ \hline & \\
$\delta(y-t), \quad t>0 $ & $ (t+z)^{-\rho}$ \\
\hline
 & \\
$y^{\nu -1}(1+y)^{-\lambda}, \quad 0 < \nu < \rho +\lambda$ &
$B(\nu, \rho + \lambda - \nu)\, z^{\nu -\rho} \,
{}_2F_1(\nu,\lambda;\rho+\lambda; 1-z)$ \\ \hline
\end{tabular}}
\caption{Pairs of functions that are related by the generalized
Stieltjes transform.}
\end{table}


\end{document}